%% file: main.tex
\documentclass[10pt,journal]{IEEEtran}

\usepackage[utf8]{inputenc}
\usepackage[T1]{fontenc}

\usepackage{graphicx}
\usepackage{physics}
\usepackage[nolist,nohyperlinks]{acronym}
\usepackage{amsmath}
\usepackage{amsthm}
\usepackage{amssymb}
\usepackage{authblk}

\usepackage{bbm}
\usepackage{bbold}
\usepackage{cite}

\usepackage[ruled,vlined,linesnumbered]{algorithm2e}
\usepackage[caption=false]{subfig}
\usepackage{graphicx}

\usepackage{tikz}
\usetikzlibrary{automata, arrows, positioning, quantikz}

\usepackage{tikz-network}
\usepackage{todonotes}
\usepackage{xcolor}

\newcommand{\hilb}[0]{%
  \mathcal{H}%
}

\newcommand{\ch}[1]{%
  \mathcal{E}_{#1}%
}

\newtheorem{theorem}{Theorem}

\newcommand{\dt}[1]{}
\newcommand{\gv}[1]{}
\newcommand{\EVM}[1]{}
\newcommand{\mg}[1]{}

\usepackage{acronym}
\newacro{BSM}{Bell state measurement}
\newacro{LLEG}{\textit{link-level entanglement generation}}
\newacro{CDF}{cumulative distribution function}
\newacro{CNOT}{controlled $X$}
\newacro{GHZ}{Greenberger-Horne-Zeillinger}
\newacro{FCFS}{first come, first serve}
\newacro{FIFO}{first-in-first-out}
\newacro{LIFO}{last-in-first-out}
\newacro{QFI}{Quantum Fisher Information}
\newacro{QFIM}{Quantum Fisher Information Matrix}
\newacro{MGF}{moment generating function}
\newacro{NISQ}{Noisy Intermediate-Scale Quantum}
\newacro{LST}{Laplace–Stieltjes transform}
\newacro{YQF}{youngest qubit first}
\newacro{OQF}{oldest qubit first}
\newacro{e2e}{end-to-end}

\title{On the Analysis of Quantum Repeater Chains with Sequential Swaps}

\author[1]{Matheus Guedes de Andrade}
\author[2]{Emily A. Van Milligen}
\author[3, 4]{Leonardo Bacciottini}
\author[1]{Aparimit Chandra}
\author[1]{Shahrooz Pouryousef}
\author[1]{Nitish K. Panigrahy}
\author[1, 5]{Gayane Vardoyan}
\author[1]{Don Towsley}

\affil[1]{Manning College of Information and Computer Sciences, University of Massachusetts Amherst}
\affil[2]{Department of Physics, The University of Arizona}
\affil[3]{Department of Information Engineering, University of Florence}
\affil[4]{Department of Information Engineering, University of Pisa}
\affil[5]{EEMCS and QuTech, Delft University of Technology}

\begin{document}

\maketitle

\input{src/abstract}
\input{src/intro}
\input{src/background}
\input{src/fidelity}
\input{src/oneShot}
\input{src/PoissonExponential}
\input{src/parameters}
\input{src/conclusion}

\newpage
\bibliographystyle{unsrt}
\bibliography{references}

\end{document}

%% file: src/abstract.tex
\begin{abstract}
We evaluate the performance of two-way quantum repeater chains with sequential entanglement swapping. Within the analysis we consider memory decoherence, gate imperfections, and imperfect link-level entanglement generation. Our main results include closed-form expressions for the average entanglement fidelity of the generated end-to-end entangled states. We generalize previous findings for the one-shot fidelity analysis and study the case where repeater chains serve end-to-end requests continuously. We provide solutions to the continuous request scenario by combining results from quantum information theory and queuing theory. Finally, we apply the formulas obtained to analyze the impacts of hardware parameters, i.e., coherence times and gate fidelity, and distance on the entanglement fidelity and secret key rate of homogeneous quantum repeater chains.
\end{abstract}

%% file: src/intro.tex
\section{Introduction}

Quantum networks promise revolutionary applications by enabling quantum processors to communicate quantum information. Quantum communication is critical for distributed quantum computing~\cite{buhrman2003distributed} and sensing~\cite{zhuang2018distributed}, and device-independent quantum key distribution~\cite{vazirani2019qkd}. Quantum network architectures are characterized as one- and two-way quantum communication networks. One-way quantum networks enable direct state transfer among network nodes, while two-way networks generate remote entanglement to be consumed by the nodes.
Two-way quantum networks have lower hardware requirements than one-way quantum networks, serving as candidates for near-term quantum networks.

There are two fundamental difficulties in the development of quantum networks. First, photons are the principal quantum information carriers and photonic communication is lossy. Attenuation in optical fibers and the atmosphere increases exponentially with distance, making long-distance quantum communication challenging. Further complicating matters, the no-cloning theorem prevents combating loss with classical replication techniques~\cite{nielsen2010quantum}.
Second, quantum information is inherently fragile. Quantum networks will demand the interconnection of a myriad of systems capable of enabling large-scale quantum communication, such as quantum memories, optical switches, and quantum interconnects~\cite{awschalom2021development, craddock2024automated}. Moreover, processing qubits with \ac{NISQ} hardware also introduces unavoidable errors~\cite{Preskill2018quantumcomputingin}.

There are different approaches to mitigate loss in quantum networks. Near-term networks combat exponential losses through multiplexing~\cite{patil2021multiplexing}, while more advanced quantum networks will utilize quantum error correction to combat photon loss~\cite{muralidharan2016optimal}. Losses make entanglement generation over network links a probabilistic process, which has a direct impact on the performance of quantum networks. Network performance is also affected by decoherence and noise: near-term networks will rely on purification or even directly utilize the remote entanglement generated by the network to serve applications, while more advanced networks will use quantum error correction for fault-tolerant quantum communication.

Assessing the performance of quantum network architectures is essential to developing efficient quantum networks. In this setting, rate, and fidelity are two key performance metrics. Rate relates to the number of transmitted qubits and generated entangled states among end-nodes in one- and two-way quantum networks, respectively. Fidelity quantifies the noise introduced by network operations during quantum communication. Computing entanglement rate and fidelity \dt{statistics} is significant since they allow one to quantify how well a quantum network serves different applications. For instance, quantum key distribution with the BB84 protocol~\cite{bennet2014quantum} tolerates any fidelity greater than 0.84~\cite{vardoyan2023quantum}, while verifiable blind quantum computation requires a minimum of 0.974 fidelity~\cite{van2024hardware}.

Quantum repeater chains are linear quantum networks that interconnect a pair of end-nodes. In the two-way case, the behavior of quantum repeater chains is determined by the ordering of entanglement swapping. Chains with sequential entanglement swapping are of interest due to their simplicity, and sequential swapping enables connectionless two-way quantum networks~\cite{lu2024connectionless}.

Analytical and numerical/simulation solutions have been developed for the analysis of quantum repeater chains, leading to progress in the quantification of fidelity and rate~\cite{brand2020efficient, coopmans2021netsquid,kamin2023exact,goodenough2024noise}.
Analytical solutions typically focus on one-shot regimes where the network is assumed to exclusively serve one communication request at time.
Repeater chains will likely be used to serve multiple requests between end-nodes, and analyzing this scenario is of practical interest.

\subsection{Contributions}

This paper provides an analytical performance analysis of two-way quantum repeater chains with sequential entanglement swapping. It adds to the literature by combining techniques from quantum information theory and performance evaluation to obtain closed-form expressions for the average fidelity of distributed quantum states. In short, the time that qubits are kept in memories throughout entanglement generation is the essential metric one needs to compute in order to calculate \ac{e2e} entanglement fidelity.
We use results from queuing theory to obtain waiting times for sequential entanglement swapping when end-nodes communicate continuously through a quantum repeater chain. Moreover, we provide a straightforward non-queuing theoretic extension to previous results that derive closed-form solutions for the one-shot entanglement fidelity of sequential entanglement swapping~\cite{kamin2023exact}.
Our detailed contributions are as follows:
\begin{itemize}
    \item We compute closed-form expressions for the average fidelity achieved by sequential entanglement swapping over a given network path in the one-shot regime. This extends previous work~\cite{kamin2023exact} by allowing nodes to have quantum memories with different coherence times and different quality quantum gates;
    \item We use queuing theory to compute closed-form expressions for the average entanglement fidelity when a repeater chain is used to generate a stream of \ac{e2e} entangled states. We model repeater chains as networks of queues in tandem, where \ac{e2e} entanglement request streams are described by Poisson processes, and link-level entanglement generation times are exponentially distributed;
    \item We apply closed-form expressions derived for average \ac{e2e} fidelity to analyze the performance of homogeneous repeater chains. We investigate the effect of different hardware parameters, e.g., coherence times and gate infidelities, and distance on the average \ac{e2e} entanglement fidelity and secret key rate.
\end{itemize}

The remainder of this paper is organized as follows. We specify the system model considered in this paper in Section~\ref{sec:background}. In Section \ref{sec:averageQuantumStates}, we provide expressions for the average fidelity of entangled states obtained from entanglement swapping. Closed-form expressions for the average fidelity in the one-shot regime and the queuing model are presented in Sections \ref{sec:oneShot} and \ref{sec:PoissonExponential}, respectively. The numerical analysis of homogeneous repeater chains's performance is reported in Section~\ref{sec:parameters}. Finally, we present concluding remarks in Section~\ref{sec:conclusion}.





%% file: src/background.tex
\section{System Model}\label{sec:background}

\begin{figure*}
    \begin{centering}
        \subfloat[Sequential entanglement swapping]{\includegraphics[height=0.47\linewidth]{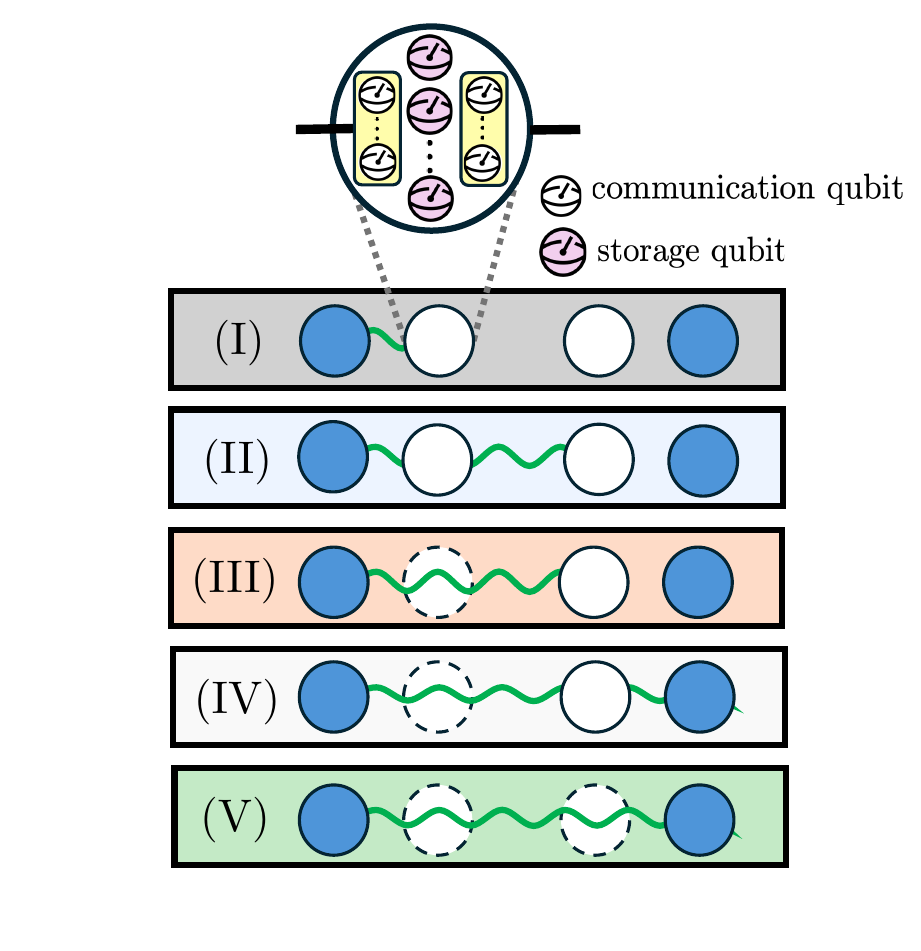} \label{subfig:seq}} \hfill
        \subfloat[Time Diagram] {\includegraphics[height=0.47\linewidth]{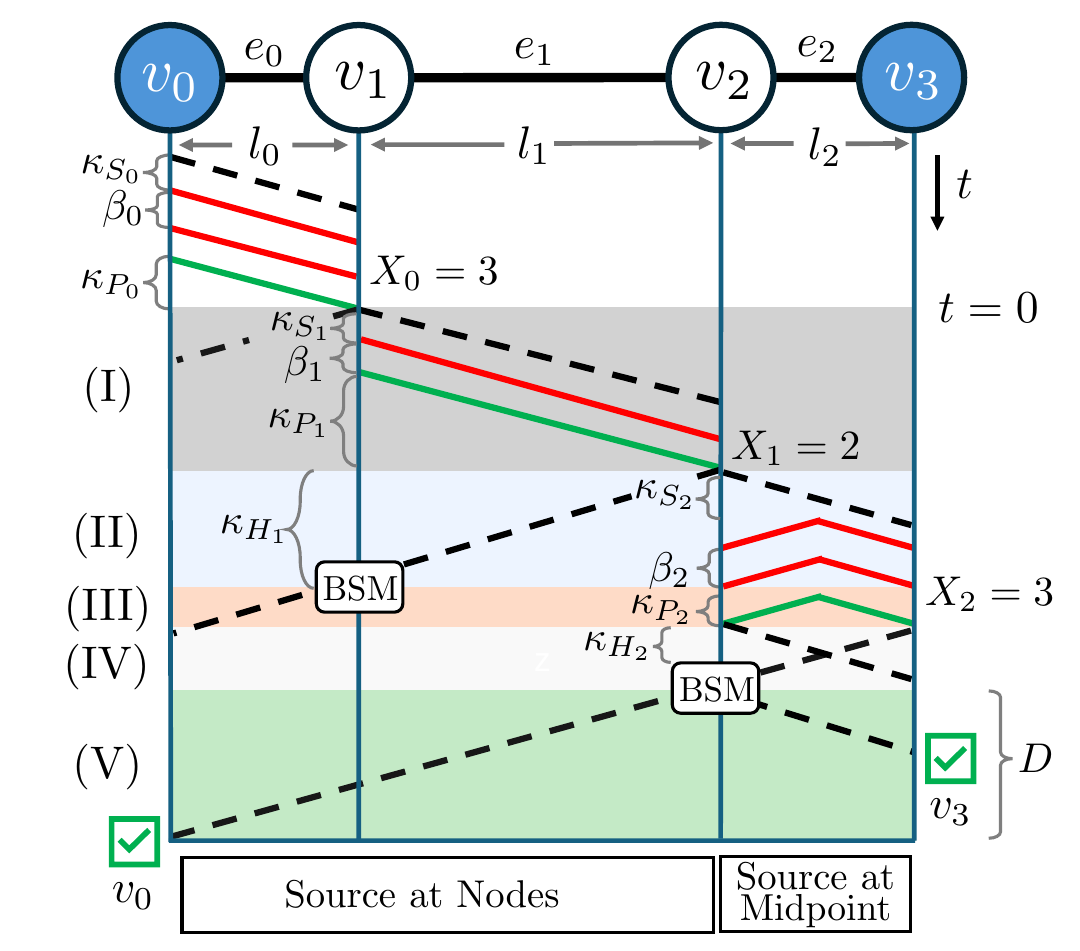}\label{subfig:timing}}
    \end{centering}
    \centering
    \caption{The blue nodes represent end nodes that wish to share a pair of entangled qubits, while the white nodes represent quantum repeaters. In (a), green wavy lines represent successfully generated entangled links. The inset shows the types of qubits in a quantum repeater considered in this paper. The different network states are numbered to show how the qubit progresses from the source to the destination. In (b), a sequence diagram is shown that shows the progression through and the time spent in each of the states numbers in (a). Dashed lines between nodes indicate classical communications. Solid lines correspond to LLEG attempts with green (red) lines correspond to the successful (failed) transmission and storage of the entangled qubits. Wait-time calculations start after the first successful link is generated at time $t=0$. End nodes do not know that the final entangled link has been distributed until after a delay $D$.}
    \label{fig:Timing}
\end{figure*}

A quantum network is a system formed by connecting quantum processors with quantum channels, thereby enabling the communication of quantum information. We denote a quantum network as a graph $G = (V, E)$, with $v\in V$ representing a quantum-equipped node and $e=(u,v)\in E$ representing a physical link between nodes $u$ and $v$, e.g., an optical fiber connection. Paths in $G$ can be considered as repeater chains and, throughout this work, we utilize $P = \{v_0, \ldots, v_{n}\} \subseteq G$ to denote a repeater chain with $n$ links. We use $l_j \in \mathbb{R}^{+}$ to denote the length of link $e_j = (v_j, v_{j + 1})$, for all $j \in \{0,\ldots, n -1\}$. From now on, we will refer to link $e_j$ as the $j$-th link of the chain or simply as link $j$.
We assume that nodes have access to unlimited matter qubits, i.e., quantum memories for storage.
\gv{Do you want to say something (or differentiate between) storage and communication qubits?}


\subsection{Link-level Entanglement Generation}

Nodes in $G$ are capable of generating entanglement with their neighbors through the quantum channels that interconnect them, a process that we hereafter refer to  as \ac{LLEG}. In principle, \ac{LLEG} on link $j$ results in the heralded generation of a Bell state, e.g., $\ket{\Phi^+} = (\ket{00} + \ket{11}) / \sqrt{2}$, held between nodes $v_{j}$ and $v_{j + 1}$.
This process is probabilistic, succeeding with probability $p_j$ which depends on $l_j$, for $j \in \{0, \ldots, n - 1\}$. In addition to its dependence on $l_j$, $p_j$ captures all hardware imperfections affecting the success probability of entanglement generation, e.g., coupling efficiencies between photons and optical cavities.
In practical scenarios, \ac{LLEG} is not capable of generating pure Bell states. We model the entangled state generated on the $j$-th link as a Werner state of form
\begin{align}
    & \Phi_{w_j}^{+} = w_j \Phi^{+} + (1 - w_j) \frac{\mathbb{1}}{4},\label{eq:WernerState}
\end{align}
where $\Phi^{+} = \dyad{\Phi^{+}}$, $\mathbb{1}: \hilb^{4} \to \hilb^{4}$ is the identity operator, and $w_j \in [0, 1]$. Werner states are a special type of Bell diagonal states, i.e., states whose eigenvectors are Bell states.

We assume for this paper that the system contains two different classes of qubits: communication qubits and storage qubits, shown in Fig.\ref{fig:Timing}. The communication qubits interface directly with the optical fiber and can be swapped with the storage qubits. In addition, communication qubits can emit photons whereas storage qubits have higher coherence times. Entanglement swapping can be performed to deterministically transfer entanglement between communication qubits and storage. 
\gv{Here it may be nice to provide some experimental examples, e.g., color centers, that operate this way, to motivate these assumptions. Also, how many interfaces are there, compared to the number of matter qubits? It may be good to provide an architecture Fig.here.}

Since \ac{LLEG} attempts are probabilistic, the time $G_{j}$ required for \ac{LLEG} to succeed on the $j$-th link is a random variable. Generation is performed in an ``attempt until success manner''. The precise description of $G_{j}$ depends on how the entanglement is generated. In general,
\begin{align}
    & G_{j} =  \kappa_{S_j} + \kappa_{P_j}+\kappa_{H_j} + \beta_j (X_j-1),\label{eq:LLEGtime1}
\end{align} 
where $\kappa_{S_j} \in \mathbb{R}^{+}$ is the classical communication time required for nodes $v_{j}$ and $v_{j + 1}$ to coordinate, $\kappa_{P_j} \in \mathbb{R}^{+}$ is the time the entangled qubits spend propagating along the optical fiber between $v_{j}$ and $v_{j + 1}$, $\kappa_{H_j} \in \mathbb{R}^{+}$ is the classical communication time required for both nodes to confirm the entangled qubits have been successfully stored at nodes $v_{j}$ and $v_{j + 1}$, $\beta_j \in \mathbb{R}^{+}$ is the time between two \ac{LLEG} attempts, and $X_j \in \mathbb{Z}$ is a geometric random variable with parameter $p_j$. The values that $\beta_j$ can take depend on the physical hardware and can be further limited by the number of communication/storage qubits at the nodes available. For this paper, we assume that nodes have access to infinite communication and storage qubits, accommodating any value of $\beta_j$. If one assumes that both $v_{j}$ and $v_{j + 1}$ do not require any form of classical coordination, or if the required classical information can be propagated simultaneously with an \ac{LLEG} attempt, then $\kappa_{S_j} = 0$. The values of $\kappa_{P_j}$ and $\kappa_{H_j}$ are determined by the entanglement generation scheme utilized. If the entanglement source is located at one of the nodes adjacent to link $j$, $\kappa_{P_j} = \kappa_{H_j}= l_j / c$, where $c$ is the speed of light in the propagation medium. On the other hand, ``source in the middle architectures'' have $\kappa_{P_j} =  l_j / (2c)$ while $\kappa_{H_j} =  l_j / c$ , assuming that the source is placed exactly in the mid-point of link $e$. For ``meet in the middle'' architectures, $\kappa_{P_j} = \kappa_{H_j} =  l_j / (2c)$.  We can rewrite \eqref{eq:LLEGtime1} as 

\begin{align}
    &G_{j} =  \kappa_{j} + \beta_j X_j,\label{eq:LLEGtime}
\end{align} 
where $\kappa_{j}=\kappa_{S_j}+\kappa_{P_j}+\kappa_{H_j}-\beta_j$ in order to isolate the random variable $X_j$. A timing diagram illustrating all of these quantities is shown in Fig. \ref{fig:Timing} (b). Two different source locations are depicted: source at the nodes (links $e_0$ and $e_1$) and source at the midpoint (link $e_2$).


\subsection{Sequential Entanglement Distribution}\label{sec:protocol}

Link-level entanglement is the building block for \ac{e2e} entanglement distribution. End-to-end entanglement is generated through repeated entanglement swapping operations, which expand the reach of entanglement. Our focus is on sequential entanglement distribution on a quantum repeater chain, illustrated in Fig.\ref{fig:Timing} (a). Suppose that node $v_0$ wants to generate an entangled state shared with node $v_{n}$.
Node $v_0$ starts \ac{LLEG} with $v_1$. Once \ac{LLEG} succeeds, corresponding to (I) in the figure, $v_1$ starts \ac{LLEG} with $v_{2}$. The entangled state shared between $v_0$ and $v_{1}$ waits in memory until \ac{LLEG} between $v_1$ and $v_2$ succeeds, corresponding to (II) in the figure. Once an entangled state is generated between $v_1$ and $v_2$, a \ac{BSM} operation is performed in $v_1$, generating an entangled state between nodes $v_0$ and $v_2$, corresponding to (III) in the figure. The process continues with every $v_{j}$ in $P$ performing LLEG with $v_{j + 1}$ followed by a \ac{BSM} that generates entanglement between $v_0$ and $v_{j + 1}$, until $(j + 1) = n$, shown in the Fig.by (V) for the case $n=4$. 


During the sequential entanglement distribution procedure, the memories in the nodes form buffers that store quantum information. Active qubits in the nodes waiting for next-hop \ac{LLEG} to succeed are viewed as communication requests.
The time required for a node to serve a communication request depends directly on its buffer occupancy. Thus, the time to serve an \ac{e2e} request depends on buffer occupancy throughout the network.
\gv{Or just on the path, if we assume static routing? Speaking of which, is this discussed?}

There are four key aspects one must specify in order to evaluate the performance of sequential entanglement swapping: 1) how \ac{e2e} communication requests are generated; 2) how \ac{LLEG} is triggered in the nodes throughout the process; 3) how \ac{BSM}s are performed when a node has multiple requests to serve; and 4) how the results of BSMs are propagated to the end nodes specifying which corrections must be applied to the resulting qubits. For 1), we consider both the one-shot regime---where the goal is to serve a single request---and streams of \ac{e2e} requests. For 2), \ac{LLEG} between nodes $v_i$ and $v_{i + 1}$ is triggered once \ac{LLEG} between nodes $v_{i - 1}$ and $v_{i}$ succeeds. For 3), we assume that requests in a node are served either on a \ac{OQF} or \ac{YQF} basis, and that \ac{BSM}s are performed as soon as a node shares two entangled states, i.e., \ac{LLEG} also triggers entanglement swapping in $v_i$ for $1 < i < n$. Finally for 4), we assume nodes immediately send the results of \ac{BSM}s to the end node that starts distribution, i.e., $v_0$.

\ac{OQF} and \ac{YQF} are time-based scheduling policies for entanglement swapping. Once \ac{LLEG} succeeds, \ac{OQF} chooses to serve the request that has been in the node for the longest time, while YQF chooses the one that has waited for the shortest. In both cases, decisions are made locally in the nodes, i.e., the timer for the age of a request starts when it arrives at the node. \ac{OQF} and \ac{YQF} policies are equivalent to first-in first-out and last-in first-out, respectively. \ac{YQF} policy maximizes fidelity over a one hop repeater chain~\cite{chandra2022scheduling}, and we investigate the effects of the two scheduling disciplines over a multi-hop repeater chain in this paper.

\subsection{Memory decoherence}\label{sec:decoherence}

Request waiting times, arising throughout the entanglement distribution process, have a direct impact on the quality of the distributed entangled states due to quantum state decoherence. We consider two models of decoherence while quibts are kept in memory: dephasing and depolarizing noise. Throughout this paper, the symbol $\circ$ is used to denote the composition of quantum channels, while $\bigcirc$ denotes the $n$-ary composition operation.

\subsubsection{Dephasing noise} Dephasing noise is a particular case of Pauli noise that represents a non-uniform contraction of the Bloch sphere~\cite{nielsen2010quantum}. The continuous-time dephasing channel assumes the form
\begin{equation}
    \mathcal{Z}_{\Gamma t}(\rho) = \frac{1 + e^{-\Gamma t}}{2} \rho + \frac{1 - e^{-\Gamma t}}{2} Z \rho Z,
\end{equation}
where $\Gamma$ and $t$ are the decoherence time and rate, respectively.
The composition of two dephasing channels is also a dephasing channel with form
\begin{align}
    \mathcal{Z}_{\Gamma_1 t_1} \circ \mathcal{Z}_{\Gamma_2 t_2} (\rho) = \frac{1 + e^{- \tau}}{2} \rho + \frac{1 - e^{-\tau}}{2} Z \rho Z,\label{eq:dephasingCat}
\end{align}
where $\tau = \Gamma_1 t_1 + \Gamma_2 t_2$.
The dephasing channel is a particular case of the single-qubit Pauli channel and it maps Bell-diagonal states into Bell-diagonal states. The action of the channel on a qubit of a Bell-diagonal state is symmetric---it yields the same result independent of which of the two qubits undergoes the depolarizing noise---and given by
\begin{align}
    & \mathbb{1} \otimes \mathcal{Z}_{\Gamma t}(\rho) = \frac{1 + e^{-\Gamma t}}{2} \rho + \frac{1 - e^{-\Gamma t}}{2} Z \rho Z,\label{eq:dephasedBellPair}
\end{align}
where $\rho$ is any Bell-diagonal state.

\subsubsection{Depolarizing noise} The depolarizing channel describes an application of isotropic noise to a quantum state, and can be interpreted as a uniform contraction of the Bloch sphere~\cite{nielsen2010quantum}. The continuous-time version of the depolarizing channel is given by
\begin{equation}
    \mathcal{D}_{\Gamma t}(\rho) = e^{-\Gamma t}\rho + (1 - e ^{-\Gamma t})\frac{\mathbb{1}}{2},
\end{equation}
where $t$ and $\Gamma$ are the depolarizing time and rate. It is possible to demonstrate that the concatenation of two depolarizing channels is itself a depolarizing channel described by a linear combination of the depolarizing parameters following
\begin{align}
    & \mathcal{D}_{\Gamma_1 t_1} \circ \mathcal{D}_{\Gamma_2 t_2}(\rho) = e^{-\tau}\rho + (1 - e ^{-\tau})\frac{\mathbb{1}}{2},\label{eq:depolarizingCat}
\end{align}
where $\tau = \Gamma_1 t_1 + \Gamma_2 t_2$. The depolarizing channel is also a particular case of the single-qubit Pauli channel and its action on a qubit of a Bell-diagonal state assumes the form
\begin{align}
    & \mathbb{1} \otimes \mathcal{D}_{\Gamma t}(\Phi) = e^{-\Gamma t} \Phi + (1 - e ^{-\Gamma t})\frac{\mathbb{1}}{4}.\label{eq:depolarizedBellPair}
\end{align}

\subsection{Noisy entanglement swapping}

Executing perfect quantum operations is extremely challenging, if not impossible without the aid of quantum error correction. Near term devices are error prone and we assume \ac{BSM}s as imperfect processes that induce depolarizing noise. Let $\mathcal{B}: \hilb^{4} \to \hilb^{2}$ denote the error-free \ac{BSM} quantum channel, where the measured pair is traced out after corrections are applied. The imperfect \ac{BSM} operation performed in node $v_{j + 1}$ is given by the channel $\mathcal{B}^{\alpha_{v}}: \hilb^{4} \to \hilb^{2}$ of form
\begin{align}
    & \mathcal{B}^{\alpha_{v}}(\rho) = \left( \mathbb{1} \otimes \mathcal{D}^{\alpha_v}\right) \circ \mathcal{B}(\rho),\label{eq:BSMNoise}
\end{align}
where $\mathcal{D}^{\alpha_v}$ is the discrete depolarizing channel \gv{Do you want to explain why depolarization is applied only to one qubit?}
\begin{align}
    & \mathcal{D}^{\alpha_v}(\rho) = \alpha_{v} \rho + (1 - \alpha_{v}) \frac{\mathbb{1}}{2},\label{eq:disrceteDepolarizing}
\end{align} 
with $\alpha_{v} \in [0, 1]$.\footnote{Depolarization is only applied to one qubit since we mainly focus on Bell-diagonal states. For such quantum states, the transpose trick allows transforming the application of two-qubit channels into a single-qubit channel applied to one of the qubits in the entangled state~\cite{wilde2013quantum}.} Note that we refer to the discrete depolarizing channel via a superscript, and the time-dependent channel via a subscript. Moreover, the properties of the discrete depolarizing channel are similar to the continuous-time case. In particular, the composition of two channels following \eqref{eq:disrceteDepolarizing} with parameters $\alpha_1, \alpha_2 \in [0, 1]$ is a discrete depolarizing channel with parameter $\alpha_1 \alpha_2$. 

\subsection{Classical Communication Delay}\label{sec:comm_delay}

Two-way quantum communication suffers a delay from propagating classical heralding signals, i.e., outcome bits of \ac{BSM}s. In particular, this delay is the time entangled states are kept in the end-nodes until \ac{BSM} results from intermediate nodes in the chain are received.
The classical communication delay $D$ is the time elapsed between the last \ac{BSM} (performed by node $v_{n-1}$) and the time at which $v_0$ and $v_n$ receive its classical results, which is illustrated in Fig.\ref{fig:Timing}(b). If we assume that all intermediate \ac{BSM} outcomes have been sent to $v_0$, $D$ assumes the form
\begin{equation}
    D = \max \left( \frac{l_{n-1}}{c}, \hspace{.2cm} \sum_{i=0}^{n-2} \frac{l_i}{c} \right) \leq \sum_{i=0}^{n-1} \frac{l_i}{c}.\label{eq:classicalDelay}
\end{equation}

%% file: src/fidelity.tex
\section{Characterizing States after Multiple Swaps}\label{sec:averageQuantumStates}

Our goal is to analyze the efficiency of entanglement distribution in quantum networks, accounting for memory decoherence, noisy gates and \ac{LLEG}, and classical communication overhead. In this section, we describe the main mathematical tool to carry out this analysis. In particular, we derive an expression for the quantum state obtained from a sequence of noisy \ac{BSM}s applied to multiple noisy entangled pairs, when each individual pair decoheres via a quantum channel. We then apply this characterization to the case of sequential entanglement distribution in a repeater chain.

\gv{I think we have to be a little careful with how general the results are claimed to be (from the above, it sounds like they are completely general).}

\subsection{Error Composition in Entanglement Swapping}

Let $\mathcal{E}_j: \hilb^{2} \to \hilb^{2}$ denote a single qubit quantum channel, for $j \in \{1, 2, 3, 4\}$. Let $\mathcal{E}^{T}$ denote the quantum channel obtained by transposing the Kraus operators of channel $\mathcal{E}$. Let $\rho_{k} \in \hilb^{4}$ denote a two-qubit state for $k \in \{1, 2\}$.
\gv{Would be good to introduce this notation in the previous section.}
\mg{We can move all the notation to section two. Maybe we can also shorten the description by abusing notation and writing channels for the Bell state as single qubit channels.}
Let $\rho = (\ch{1}\otimes \ch{2}) ( \rho_1) \otimes (\ch{3} \otimes \ch{4}) (\rho_2)$. If $\rho_1$ and $\rho_2$ are any of the four Bell states, the transpose trick~\cite{wilde2013quantum} allows rewriting $\rho$ as the state
\begin{align}
    & \rho = ((\ch{1} \circ \ch{2}^{T})) \otimes \mathbb{1}) (\rho_1) \otimes (\mathbb{1} \otimes (\ch{4} \circ \ch{3}^{T})) (\rho_2).\label{eq:BellDiagonal}
\end{align}
A one-qubit quantum channel $\mathcal{E}: \hilb^{2} \to \hilb^{2}$ is teleportation-covariant~\cite{pirandola2017fundamental} if, for any Pauli unitary $U \in \{\mathbb{1}, X, Y, Z\}$, the equation
\begin{align}
    \mathcal{E}(U \sigma U^{\dagger}) = V \mathcal{E} (\sigma) V^{\dagger}
\end{align}
holds for a unitary $V: \hilb^{2} \to \hilb^{2}$, for any qubit state $\sigma: \hilb^{2} \to \hilb^{2}$. If the composite channels are teleportation covariant, a \ac{BSM} applied to qubits 2 and 3 in \eqref{eq:BellDiagonal} yields
\begin{align}
    & \mathcal{B}(\rho) = (\ch{1} \circ \ch{2}^{T}) \otimes (\ch{4} \circ \ch{3}^{T}) (\mathcal{B}(\rho_1 \otimes \rho_2)). \label{eq:bsm1}
\end{align}
The result of the \text{BSM} operation in the r.h.s of equation \eqref{eq:bsm1} is also a Bell state, given that both $\rho_1$ and $\rho_2$ are Bell states. Thus, the transpose trick allows further simplifying \eqref{eq:bsm1} to
\begin{align}
    & \mathcal{B}(\rho) = \mathbb{1} \otimes (\ch{4} \circ \ch{3}^{T} \circ \ch{2} \circ \ch{1}^{T})(\mathcal{B}(\rho_1 \otimes \rho_2)).\label{eq:errorConcatenation}
\end{align}

 \gv{If I apply your notation above literally, then I get $\mathbf{1}\otimes(\mathcal{E}_1\circ\mathcal{E}_2)(\Phi^+)$, i.e., without the transpose operations.}
Hence, we can model the effects of teleportation-covariant channels on entanglement swapping as a single-qubit channel applied at the end of the \ac{BSM}.
Under the assumption of Pauli noise, i.e., $\mathcal{E}_j$ is a Pauli operator for all $j$, the technique generalizes to an arbitrary number of swaps despite the swap order used, yielding the following theorem.
\begin{theorem}\label{th:channelCat}
    Let $\rho = \bigotimes_{j = 1}^{n}\Phi^{+}$ denote $n$ copies of $\Phi^{+}$. Let $\mathcal{P}_j$ denote a single-qubit Pauli channel applied to one qubit of the $j$-th Bell pair in $\rho$. The final state $\rho_{BSM}$ obtained by $(n - 1)$ swaps is
    \begin{align}
        & \rho_{BSM} = \mathbb{1} \otimes (\bigcirc_{j = 1}^{n} \mathcal{P}_j)(\Phi^{+}),\label{eq:th1}
    \end{align}
    where $\bigcirc$ denotes the $n$-ary composition operation for quantum channels.
\end{theorem}
\gv{Is $\bigcirc$ meant to be composition? If so, I think it should be mentioned here to avoid confusion.}
\begin{proof}
    Pauli channels are teleportation-covariant channels for which $\mathcal{P}^{T} = \mathcal{P}$. Since Pauli channels commute, $\mathcal{P}_j \circ \mathcal{P}_k = \mathcal{P}_k \circ \mathcal{P}_j$ for all $j \neq k$. Moreover, $\mathcal{P}_j \circ \mathcal{P}_k$ is also a Pauli channel and, thus, teleportation covariant.
    The result follows from the application of \eqref{eq:errorConcatenation} consecutively for the channels $\mathcal{P}_j$, for $j \in \{0, 1, \ldots, n - 1\}$. The ordering from 1 to $n$ in the $n$-ary concatenation operator follows from the commutative property of Pauli channels. Note that the commutative property of Pauli channels also ensures that \eqref{eq:th1} is applicable to any swap order utilized.
\end{proof}

\mg{We can improve how this is expressed for next week. For now, I 'll leave text as is so that we have an initial document to submit.}
\subsection{Closed-form Expressions for Distributed States}

We assume that imperfect \ac{BSM}s, noisy \ac{LLEG}, and quantum memory decoherence introduce Pauli errors. 
Thus, Theorem~\ref{th:channelCat} applies to the analysis of entangled states generated from a sequence of entanglement swaps in a repeater chain. Since Pauli channels commute, Theorem~\ref{th:channelCat} allows us to separate the effect of memory decoherence---which depends on the time that a qubit resides in memory---from the time-independent errors accrued from \ac{BSM}s and \ac{LLEG}.

\subsubsection{Noise in Swaps and \ac{LLEG}}

We start by analyzing the impact of noisy \ac{BSM}s and noisy \ac{LLEG}. We model noisy \ac{BSM} operations via depolarizing noise, as shown in \eqref{eq:BSMNoise}. We obtain Werner states, \eqref{eq:WernerState}, by applying a depolarizing channel $\mathcal{D}^{w_j}$ to one qubit of the pure Bell state $\ket{\Phi^{+}}$.
Entanglement swapping through a path $P$ in $G$ utilizes one Werner state generated from each link in $P$, as well as a \ac{BSM} in each intermediate node.

%
Thus, all errors from \ac{LLEG} and \ac{BSM}s are expressed by the composed channel
$\left( \bigcirc_{j = 1}^{n - 1} \mathcal{D}^{\alpha_j} \right) \circ \left( \bigcirc_{j = 0}^{n - 1} \mathcal{D}^{w_j} \right)$. The composition property of depolarizing channels, \eqref{eq:depolarizingCat}, implies that the resulting channel is a depolarizing channel $\mathcal{D}^\chi$, with $\chi$ given by
\begin{align}
    & \chi = \prod_{j = 1}^{n - 1}\alpha_j \prod_{k = 0}^{n}w_k.\label{eq:chi} 
\end{align}
Note that \eqref{eq:chi} is invariant to the swap order.

\subsubsection{Memory Decoherence and Final Expressions} We now analyze the effects of memory decoherence, and use the previous results to describe the final distributed state $\rho$. We start with the dephasing model.

The sequential entanglement swapping protocol described in Section \ref{sec:protocol} performs \ac{BSM}s as soon as possible. Hence, only one pair of qubits decoheres at any point in time during generation of end-to-end Bell pair.
Let $T_j$ denote the time that the entangled pair between nodes $v_0$ and $v_j$ decoheres, for $j = 1, \ldots, n$. Let $\Gamma_j'$ denote the dephasing parameter for the $j$-th entangled pair, for $j \in \{1, \ldots, n\}$. From Theorem~\ref{th:channelCat}, the effect of memory decoherence is captured by the composed channel $(\mathbb{1} \otimes \bigcirc_{j = 1}^{n - 1} \mathcal{Z}_{\Gamma_j' T_j})$. Classical communication delay adds decoherence at the end-nodes, that is captured by the dephasing channel $\mathcal{Z}_{\Gamma_{n}' D}$. Hence, the final entangled state takes the form
\begin{align}
    \rho = (\mathbb{1} \otimes \mathcal{D}^{\chi}) \circ (\mathbb{1} \otimes \mathcal{Z}_{\Gamma_{n}'D}) \circ (\mathbb{1} \otimes \mathcal{Z}_{\sum_{j = 1}^{n - 1} \Gamma_j' T_j})  (\Phi^{+}),\label{eq:depolarizedSwapState}
\end{align}
where $\chi$ and $D$ follow \eqref{eq:chi} and \eqref{eq:classicalDelay}, respectively.
Thus, taking
\begin{align}
    & \Delta = \Gamma_{n}'D\label{eq:Delta}
\end{align}
and
\begin{align}
    & \tau = \sum_{j = 1}^{n - 1} \Gamma_{j}' T_j\label{eq:tau}
\end{align}
allows re-rewriting \eqref{eq:depolarizedSwapState} as
\begin{align}
    & \rho = (\mathbb{1} \otimes \mathcal{D}^{\chi}) \circ (\mathbb{1} \otimes \mathcal{Z}_{\Delta + \tau})(\Phi^{+}).\label{eq:simplifiedDepolarizedSwapState}
\end{align}
The actions of dephasing and depolarizing together lead to a final state $\rho$ of the form
\begin{align}
     & \rho = \chi\left( \frac{1 + e^{-(\Delta + \tau)}}{2} \Phi^{+}
         + \frac{1 - e^{-(\Delta + \tau)}}{2}  \Phi^{-}
         \right)
         + (1 - \chi) \frac{\mathbb{1}}{4}.\label{eq:dephasedCatBellPair}
\end{align}

The case for depolarizing noise is analogous to dephasing. In this scenario, the composed action of the noise channels follows
\begin{align}
    \rho = (\mathbb{1} \otimes \mathcal{D}^{\chi}) \circ (\mathbb{1} \otimes \mathcal{D}_{\Delta}) \circ (\mathbb{1} \otimes \mathcal{D}_{\tau})  (\Phi^{+}).\label{eq:depolarizedSwapState}
\end{align}
Thus, the final state is
\begin{align}
    \rho = \chi e^{-(\Delta+ \tau)} \Phi^{+} + \left(1 - \chi e^{-(\Delta + \tau)} \right) \frac{\mathbb{1}}{4}.\label{eq:depolarizedCatBellPair}
\end{align}

\subsection{Expected Entanglement Fidelity}

We now utilize \eqref{eq:dephasedCatBellPair} and \eqref{eq:depolarizedCatBellPair} to compute the expected entanglement fidelity when decoherence times are independent random variables, i.e., $T_j$ in \eqref{eq:tau} is a random variable for all $j$ and $T_j$ is independent of $T_k$ for $j \neq k$. The only source of randomness in the aforementioned equations is $\tau$. Hence, 
the entanglement fidelity of the final state $\rho$ for the case of dephasing and depolarizing noise is
\begin{align}
    & F_{\mathcal{Z}} = \chi \frac{1 + e^{-\Delta} \mathcal{L}_\tau(1)}{2} + \frac{1 - \chi}{4},\label{eq:averageDephState}
\end{align}
and
\begin{align}
    & F_{\mathcal{D}} = \frac{1 + 3 \chi e^{-\Delta}\mathcal{L}_\tau(1)}{4},\label{eq:averageDepoState}
\end{align}
respectively, where $\mathcal{L}_\tau(x) = \mathbb{E}[e^{-x\tau}]$ is the \ac{LST} for $\tau$. Since $\tau$ is a linear combination of independent random variables, its \ac{LST} takes the form
\begin{align}
   \mathcal{L}_{\tau}(x) = \prod_{j=1}^{n - 1}\mathcal{L}_{T_j}(\Gamma_j x).\label{eq:momentProdForm}
\end{align}
The product form of the \ac{LST} facilitates the analysis since we can independently compute the \ac{LST}s for $T_j$, for $j \in \{1,\ldots, n - 1\}$, and take their product. Furthermore, \eqref{eq:averageDephState} and \eqref{eq:averageDepoState} can be utilized to analyze any swap schedules, while \eqref{eq:momentProdForm} is only applicable to schedules leading to independent waiting times.

%% file: src/oneShot.tex
\section{One-shot Analysis}\label{sec:oneShot} 

We now apply the analytical tools described in Section~\ref{sec:averageQuantumStates} to a one-shot analysis of sequential entanglement distribution. In the one-shot regime, it is assumed that only one request, i.e., one entangled pair, is to be served between nodes $v_0$ and $v_n$. Assume the quantum memories in node $v_j$ have decoherence parameter $\Gamma_j$. Node $v_0$ always stores one of the qubits of the entangled state throughout sequential generation. Therefore, entangled states shared between nodes $v_{0}$ and $v_{j}$ wait for time $T_{j}$ decohering at a rate $\Gamma_{j}' = (\Gamma_{0} + \Gamma_{k})$. Moreover, nodes have no requests stored in their buffers other than the one being served. Thus, the decoherence time $T_j$ experienced by the entangled pair shared between nodes $v_0$ and $v_j$ is the \ac{LLEG} time $G_{j}$ given by \eqref{eq:LLEGtime}.

From \eqref{eq:averageDephState} and \eqref{eq:averageDepoState}, the average entanglement fidelity is computed from the \ac{LST}s of the random variables $T_j = G_j$, for $1 < j < n - 1$. Note that the time it takes to generate entanglement between $v_0$ and $v_1$ does not contribute to decoherence, as depicted in the time diagram shown in Fig.~\ref{fig:Timing}.
Since $T_j$ follows \eqref{eq:LLEGtime}, its \ac{LST} assumes the form
\begin{align}
    & \mathcal{L}_{T_j}(x) = e^{-\kappa_j x} \frac{p_j e^{ -\beta_j x}}{1 - (1 - p_j)e^{-\beta_j x}}, \label{eq:linkMGF}
\end{align}
where the term $e^{-\kappa_j x}$ appears due to the time delay $\kappa_j$. The \ac{LST} for $\tau$ is directly obtained from \eqref{eq:momentProdForm} and \eqref{eq:linkMGF}, yielding 
\begin{align}
    & \mathcal{L}_{\tau}(x) = e^{-x(\sum_{j} \Gamma_{j}^{'} \kappa_j)} \prod_{j = 1}^{n - 1} \left( \frac{p_j e^{-x\Gamma_{j}^{'} \beta_j}}{1 - (1 - p_j)e^{-x \Gamma_{j}^{'} \beta_j}} \right)\label{eq:oneShotMGF}.
\end{align}
Note that \eqref{eq:oneShotMGF} applies to both dephasing and depolarizing decoherence models. Closed-form expressions for the average fidelity are obtained by substituting \eqref{eq:oneShotMGF} into \eqref{eq:averageDephState} and \eqref{eq:averageDepoState}.

%% file: src/PoissonExponential.tex
\section{Repeater Chains under Poisson Arrivals and Exponential Service}\label{sec:PoissonExponential}
\begin{figure}
\centering
\includegraphics[scale=.25]{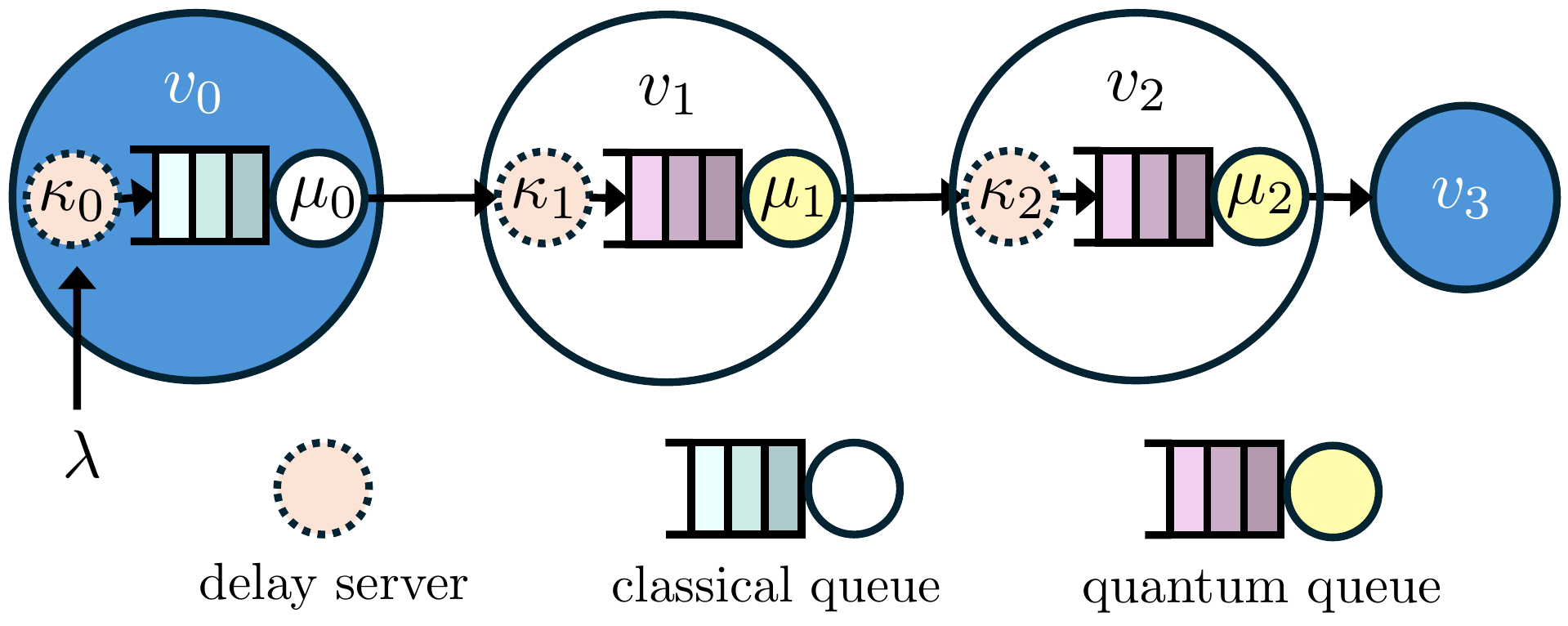}
    \caption{M/M/1 queue system with fixed time delays. Sequential entanglement generation maps to a system of M/M/1 queues in tandem, where $\lambda$ is the rate of the Poisson arrival process in $v_0$, $\mu_j$ is the exponential service rate of link $j$, and $\kappa_j$ is a deterministic fixed time delay. }
    \label{fig:Queue}
\end{figure}

The one-shot analysis presented in Section~\ref{sec:oneShot} serves as an upper bound on average fidelity for realistic scenarios, since buffers in the nodes are assumed to be empty and end-nodes generate a single entangled pair between each other. It is of practical interest to study the case where an end-node generates a stream of entanglement requests and nodes store requests to be served. In this section, we evaluate the performance of a repeater chain under two assumptions: 1) node $v_0$ generates a stream of requests targeting node $v_{n}$ following a Poisson arrival process with rate $\lambda$; and 2) the \ac{LLEG} time between nodes $v_j$ and $v_{j + 1}$ can be modeled as the concatenation of an exponential random variable with parameter $\mu_j$ and a fixed delay. Our analysis yields closed-form solutions for $\mathcal{L}_{\tau}$ when queues in the network are operated under both \ac{OQF} and \ac{YQF}. Once more, closed-form solutions for the average fidelity are obtained by substituting $\mathcal{L}_{\tau}$
into \eqref{eq:averageDephState} and \eqref{eq:averageDepoState}.

\subsection{Queues in Tandem}
Node $v_0$ is continuously generating requests, and we model the repeater chain $P = \{v_0, \ldots, v_{n}\}$ as a sequence of $n$ M/M/1 queues \cite{harchol2013performance} in tandem with additional delays, as shown in Figure~\ref{fig:Queue}. 
In this setting, the time $T_j$ that a qubit decoheres in node $j$ is given by
\begin{align}
    & T_j = S_j + \kappa_j,
\end{align}
where $S_j$ is the sojourn time, i.e., the time that a request spends in a queue from its arrival until its departure, for the $j$-th queue (node $v_{j}$), and $\kappa_j$ is the fixed delay shown in~\eqref{eq:LLEGtime}.

M/M/1 queues in tandem have been extensively studied in the literature of queuing theory, and sojourn times have closed-form~\cite{harchol2013performance}. We apply Burke's theorem: each queue in the chain behaves as an independent M/M/1 queue, with Poisson arrival process with rate $\lambda$ and exponential service time $\mu_j$~\cite{harchol2013performance}. Thus, sojourn time $S_j$ is independent of $S_k$, for $k \neq j$. Note that the delay servers do no break the properties of the queues in tandem provided by Burke's theorem.


\subsection{Exponential Model for Link-level Entanglement Generation}
Waiting times $L_j$, $j\in\{0,\ldots,n-1\}$ for ``attempt until success" \ac{LLEG} methods follow \eqref{eq:LLEGtime}. We now describe how to choose parameters to model the \ac{LLEG} time as the concatenation of an exponential random variable and a constant delay. In particular, we model $G_j$ as 
\begin{align}
    & G_j = \kappa_j + Y_j,
\end{align}
where $Y_j$ is an exponential random variable with mean $\mu_j$, for all $j \in \{0, 1, \ldots, n-1\}$.
In this setting, the goal is to find $\mu_j$ such that the exponential random variable $Y_{j}$ captures the random component of the \ac{LLEG} process through link $j$, i.e., $Y_{j}$ captures the random variable $\beta_j X_j$ in \eqref{eq:LLEGtime} for all $j \in \{0, 1, \ldots, n - 1\}$.

We now describe two different choices of $\mu_j$ that provide an upper bound and an approximation on the average fidelity of entanglement. Let $C_{X}$ denote the \ac{CDF} of the random variable $X$. For an upper bound, we select $\mu_j$ by setting the \ac{CDF} of $Y_{j}$ at times $k \beta_j$ to be equal to the CDF of $X_j$ evaluated at $k$, i.e., $C_{Y_{j}}(k \beta_j) = C_{X_j}(k)$ for $k \in \mathbb{Z^+}$.
Specifically, we select $\mu_j$ such that
\begin{align}
    1 - e^{\mu_j \beta k} = 1 - (1 - p_j)^{k} \hspace{.2cm} \forall k \in \mathbb{Z^+}, \label{eq:upperBoundCondition}
\end{align}
which yields
\begin{align}
    & \mu_j = \frac{-\log(1 - p_j)}{\beta_j}.\label{eq:exponentialUpperBound}
\end{align}
The approximation is obtained by selecting
\begin{align}
    & \mu_j = \frac{\beta_j}{p_j},\label{eq:exponentialApprox}
\end{align}
such that the expected values of $\beta_j X_j$ and $Y_j$ are the same. The error of this approximation decreases as $p_j$ becomes small.


\subsection{Decoherence time LST for OQF} \dt{introduce FIFO or reintroduce}

When requests are served at each node with \ac{OQF} scheduling policy, $S_j$ has probability density function given by $f(x) = (\mu_j - \lambda)e^{-(\mu_j - \lambda)x}$, for $t > 0$. Hence, the \ac{LST} of $T_j$ has the form
\begin{align}
    & \mathcal{L}_{T_j}(x) = e^{-\kappa_j x} \left( \frac{\mu_j - \lambda}{\mu_j - \lambda + x} \right).
\end{align}
Thus, the LST of $\tau$ is given by
\begin{align}
    & \mathcal{L}_{\tau}(x) = \prod_{j = 1}^{n - 1} e^{-\kappa_j \Gamma_j' x} \left( \frac{\mu_j - \lambda}{\mu_j - \lambda + x\Gamma_j^{'}} \right).\label{eq:queueGenericMGF}
\end{align}
Combining the last equation with \eqref{eq:exponentialUpperBound} and \eqref{eq:exponentialApprox} yields
\begin{align}
    & \mathcal{L}_{\tau}(x) = \prod_{j = 1}^{n - 1} e^{-\kappa_j \Gamma_j' x} \left( \frac{\log(1 - p_j) + \beta_j \lambda}{\log(1 - p_j) + \beta_j \lambda - x \beta_j \Gamma_j^{'}} \right)\label{eq:queueMGF}
\end{align}
and
\begin{align}
    & \mathcal{L}_{\tau}(x) = \prod_{j = 1}^{n - 1} e^{-\kappa_j \Gamma_j' x} \left( \frac{\beta_j - p_j \lambda}{\beta_j - \lambda p_j + x p_j \Gamma_j^{'}} \right).\label{eq:queueMGFApprox}
\end{align}

\subsection{Decoherence time LST for YQF}

The \ac{LST} of the M/M/1 queue sojourn time distribution operating with YQF
also has closed-form. In particular, it is the \ac{LST} of the busy period distribution of the M/M/1 queue. 
The busy period is the time it takes for the queue to leave the empty state, i.e., a job arrives and the queue is empty, and return to the empty state, i.e., a job departs and the queue is empty.
The busy period $B_j$ of the $j$-th queue has closed-form \ac{LST}~\cite{wishart1960queuing} given by 
\begin{align}
    & \mathcal{L}_{B_j}(x) = \frac{(\mu_j + \lambda + x) - \sqrt{x^2 + 2 x(\lambda + \mu_j) + (\lambda - \mu_j)^2)}}{2 \lambda}.\label{eq:busyMGF}
\end{align}
Since $S_j = B_j$ and $T_j = S_j + \kappa_j$, \eqref{eq:momentProdForm} yields:
\begin{align}
    & \mathcal{L}_{\tau}(x) = \prod_{j = 1}^{n - 1} e^{-\kappa_j \Gamma_j' x} \mathcal{L}_{B_j}(x\Gamma_j').\label{eq:LIFOMGF}
\end{align}
For the sake of space, we do not substitute \eqref{eq:exponentialUpperBound}, \eqref{eq:exponentialApprox}, and \eqref{eq:busyMGF} into \eqref{eq:LIFOMGF}.

%% file: src/parameters.tex
\section{Performance evaluation of homogeneous repeater chains}\label{sec:parameters}

\begin{figure}
    \begin{centering}
        \subfloat[Average Fidelity]{\includegraphics[width=\linewidth]{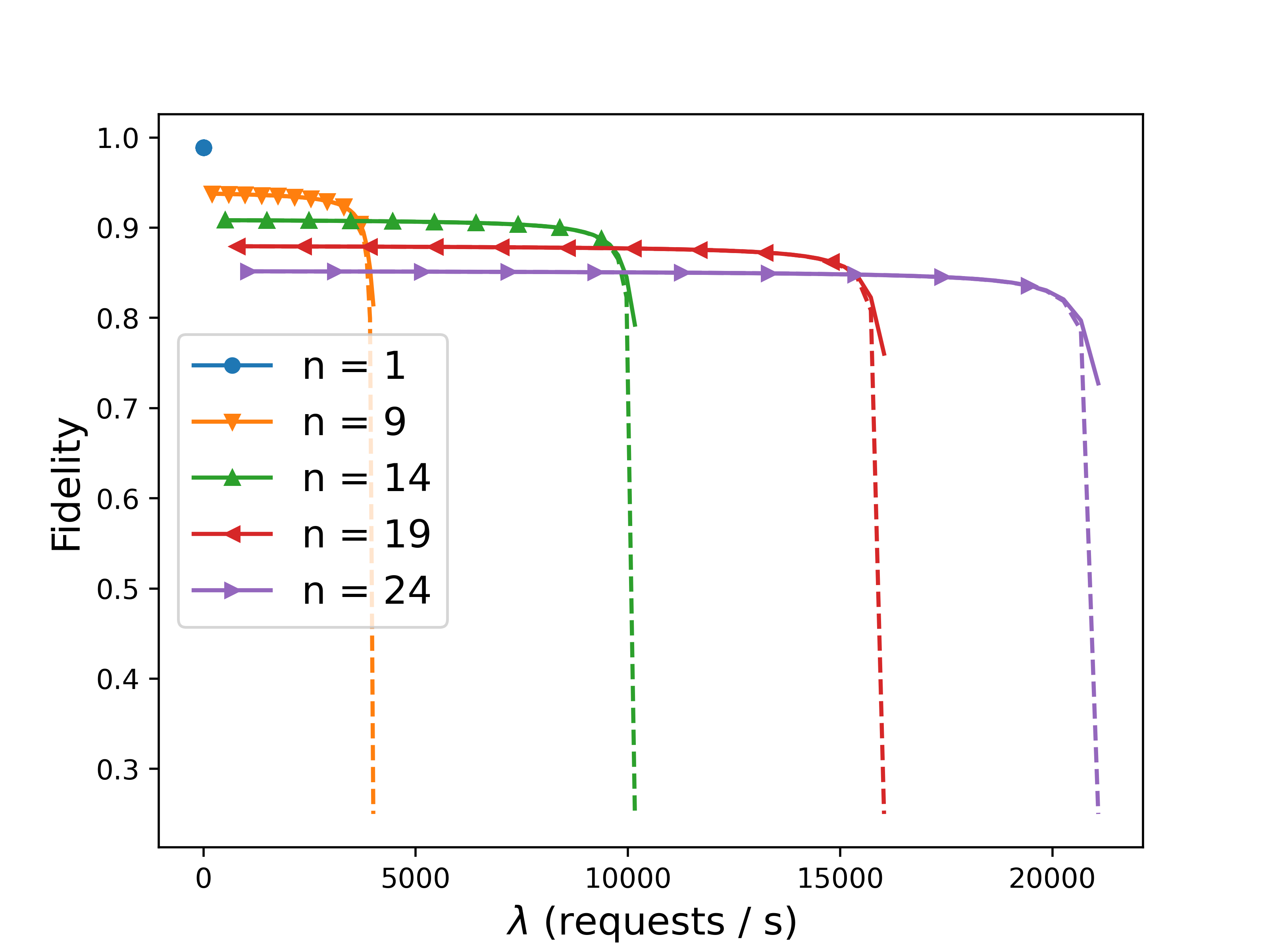} \label{subfig:fidDeph}} \hfill
        \subfloat[Average SKR] {\includegraphics[width=\linewidth]{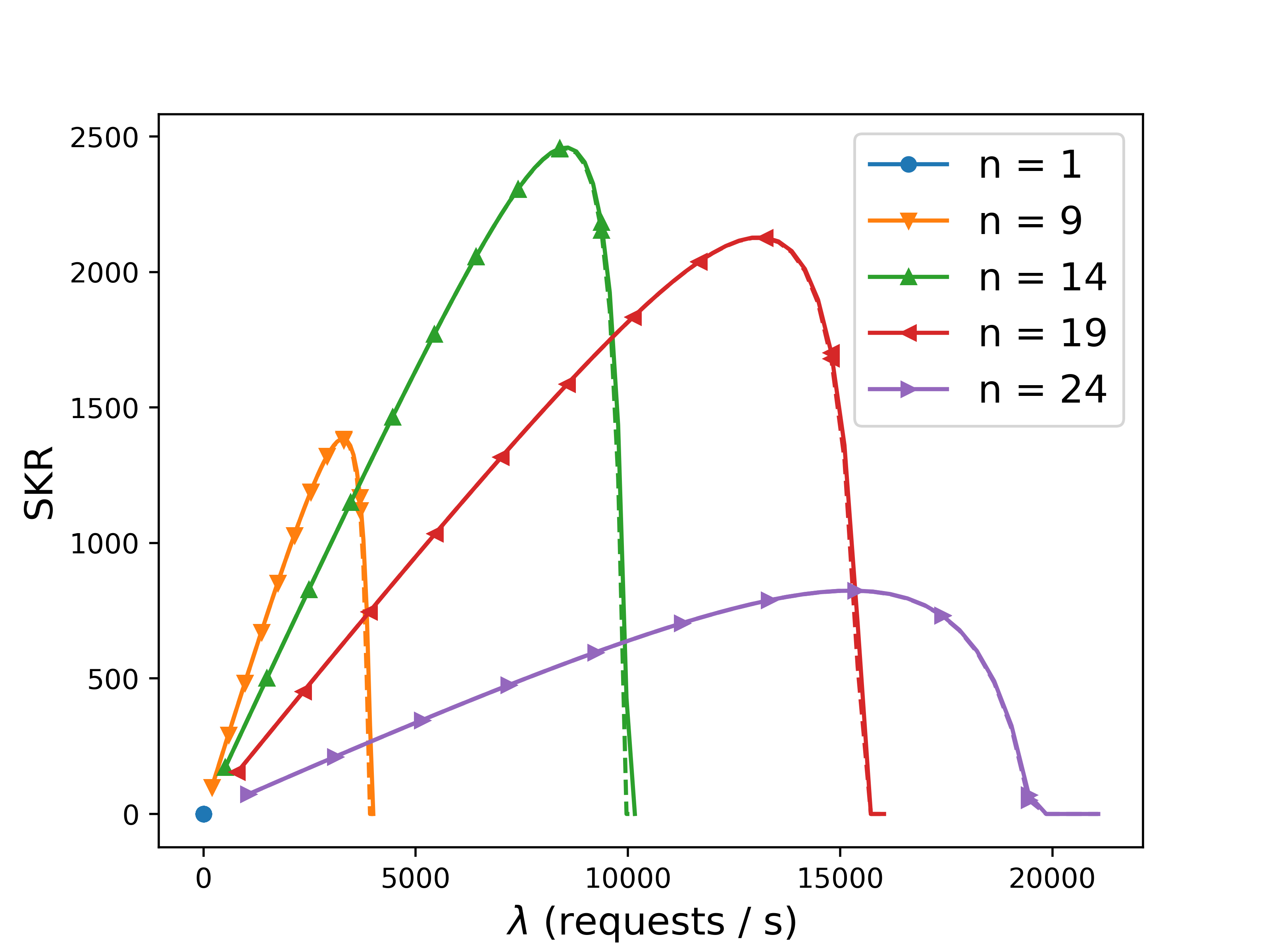}\label{subfig:skrDeph}}
    \end{centering}
    \caption{Fidelity and SKR versus request arrival rate for homogeneous repeater chains. Solid and dashed lines represent queues with YQF  and OQF policies, respectively. The total distance between repeaters is $l = 500$ km. The inter-repeater distance for each curve is $l / (n - 1)$ km. $w^{*} = 0.995$, $\alpha^{*} = 0.996$, and $T^{*} = 1$s.}
    \label{fig:fidelity}
\end{figure}

In this section, we apply the closed-form expressions obtained in Section \ref{sec:PoissonExponential} to evaluate the performance of homogeneous repeater chains. We restrict ourselves to the analysis of depolarizing noise, utilizing \eqref{eq:averageDepoState} to calculate average fidelity. The analysis shown here can be easily replicated to the depolarizing case by utilizing \eqref{eq:averageDephState} instead of \eqref{eq:averageDepoState}. We focus on upper bounds in performance, utilizing $\mu_j$ given in \eqref{eq:exponentialUpperBound} throughout our evaluation. In the homogeneous setting, repeaters are equidistant and have the same hardware parameters, i.e., $\Gamma_j = \Gamma^{*}$ and $\alpha_j = \alpha^{*}$, for $j \in \{0, 1, \ldots, n\}$. Moreover, \ac{LLEG} on every link in the chain is also characterized by the same parameter, such that $w_{j} = w^{*}$ for all $j\in \{0, 1,\ldots, n - 1\}$.
Throughout evaluation, we select $\beta_j = 10^{-5}s$, and assume that \ac{LLEG} success probability is given by
\begin{align}
    p_j = (0.7)^{2} \times e^{l_j / {22}},
\end{align}
where 0.7 denotes the emission/collection photon efficiency, and the coefficient of $22^{-1}$ in the exponential argument is compatible with optical fibers with attenuation coefficient of $0.2$ dB / km. Memory decoherence is analyzed with respect to coherence times such that $\Gamma^{*} = 1 / T^{*}$ with $T^{*}$ given in seconds.

In addition to average fidelity, we compute the average secret key rate (SKR) for the BB84~\cite{bennet2014quantum} protocol. The avergage SKR for Werner states assumes the form
\begin{align}
    & \text{SKR} = \max \left( 0,  \lambda \left( 1 - 2 H_2 \left( \frac{2(1 - \mathbb{E}[F])}{3} \right) \right) \right),\label{eq:skr}  
\end{align}
where $H_2$ is the binary entropy function~\cite{vardoyan2023quantum}.

\subsection{Fidelity and Secret Key Rate vs Request Rate}

Our results for average fidelity and SKR with $\lambda$ are reported in Figure~\ref{fig:fidelity}. We analyze chains with different numbers of repeaters, while maintaining a fixed \ac{e2e} distance of 500km. The curves show that average fidelity depends both on the number of repeaters and the rate of requests served by system. In particular, fidelity decreases monotonically with $\lambda$ since increasing the rates leads to increases in waiting times in the queues. Additionally, fidelity also decays monotonically with the number of repeaters in the chain for the parameter values considered. This stems the fact that adding repeaters increases both the number of imperfect \ac{BSM}s performed and the number of imperfect entangled states used for \ac{e2e} distribution.

Fidelities drop significantly when $\lambda \geq \mu$. In this case, the system of queues in tandem is not stable and wait times for OQF scheduling approach infinity. YQF still achieves meaningful fidelities when $\lambda$ approaches $\mu$ since newly generated EPR pairs are used to serve requests.

The relationship between SKR and $\lambda$ differs from that between fidelity and $\lambda$. SKR depends on the fidelity-rate trade-off. SKR increases from an increase in $\lambda$ up to a certain point after which it decreases due to a decrease in fidelity. Furthermore, the number of repeaters that maximize the SKR depends on the desired rate, i.e., the curve that attains the maximum SKR in Fig.\ref{subfig:skrDeph} depends on $\lambda$.

Instability in the system also impacts secrete key rate. Differently from fidelity, SKR approaches zero under both OQF and YQF. This behavior stems from the fact that the binary entropy shown in \eqref{eq:skr} exhibits a threshold behavior with fidelity. Once the system approaches instability, fidelity for both OQF and YQF drops below this threshold yielding an SKR of zero.

Queues operating in \ac{YQF} (solid lines) have higher \ac{e2e} fidelities than \ac{OQF} (dahsed lines). This behavior has been initially reported for the average fidelity of teleportation over a single link~\cite{chandra2022scheduling}, and it is corroborated by our experiments for the case of a repeater chain. From now on, we focus on analyzing repeater chains with YQF, given its superior performance.

\subsection{Optimal Secret Key Rate vs Distance}

\begin{figure}
    \begin{centering}
        \subfloat[$\lambda = 2000$ requests/s]{\includegraphics[width=\linewidth]{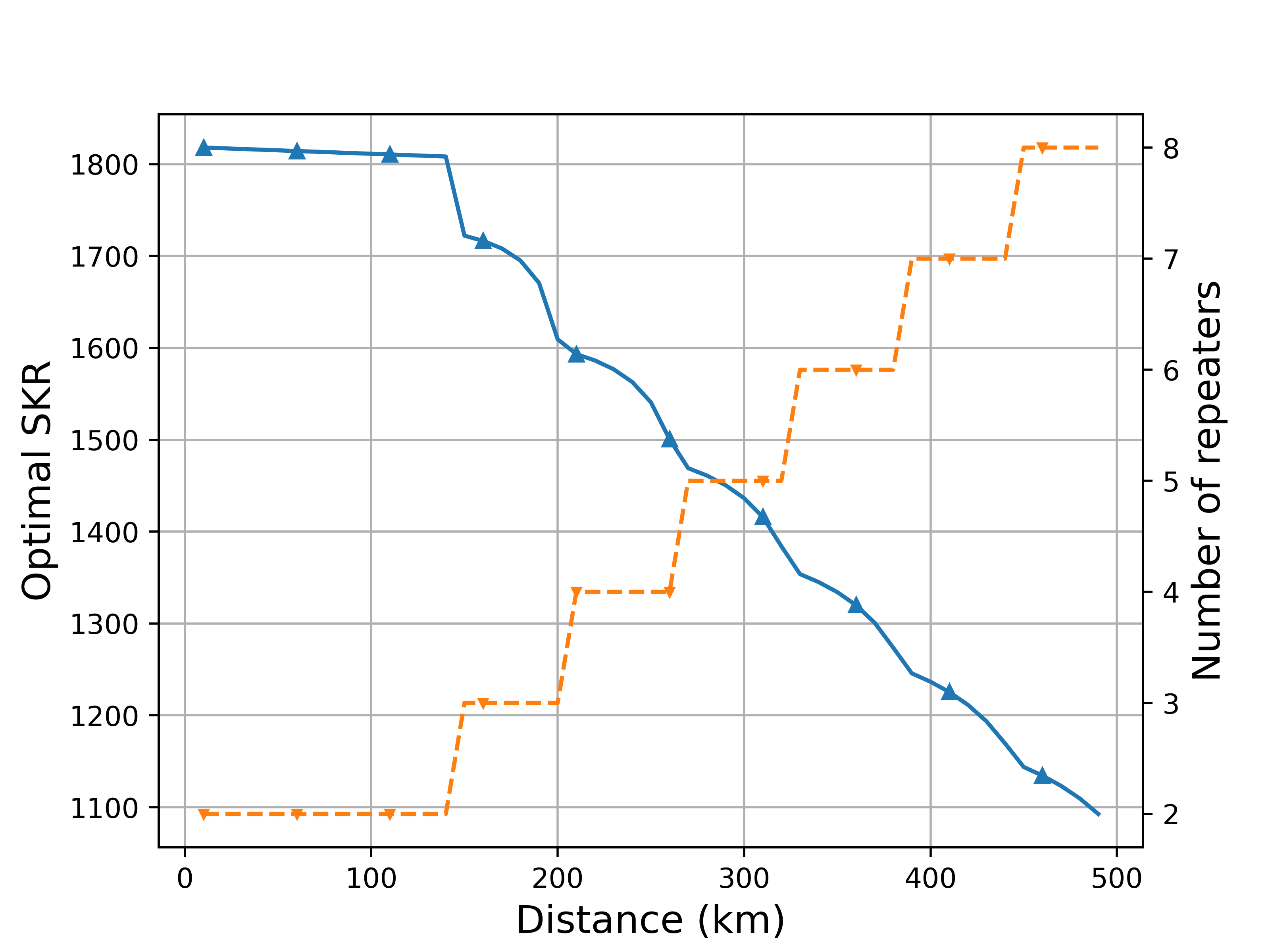} \label{subfig:lowLoad}} \hfill
        \subfloat[$\lambda = 8000$ requests/s]{\includegraphics[width=\linewidth]{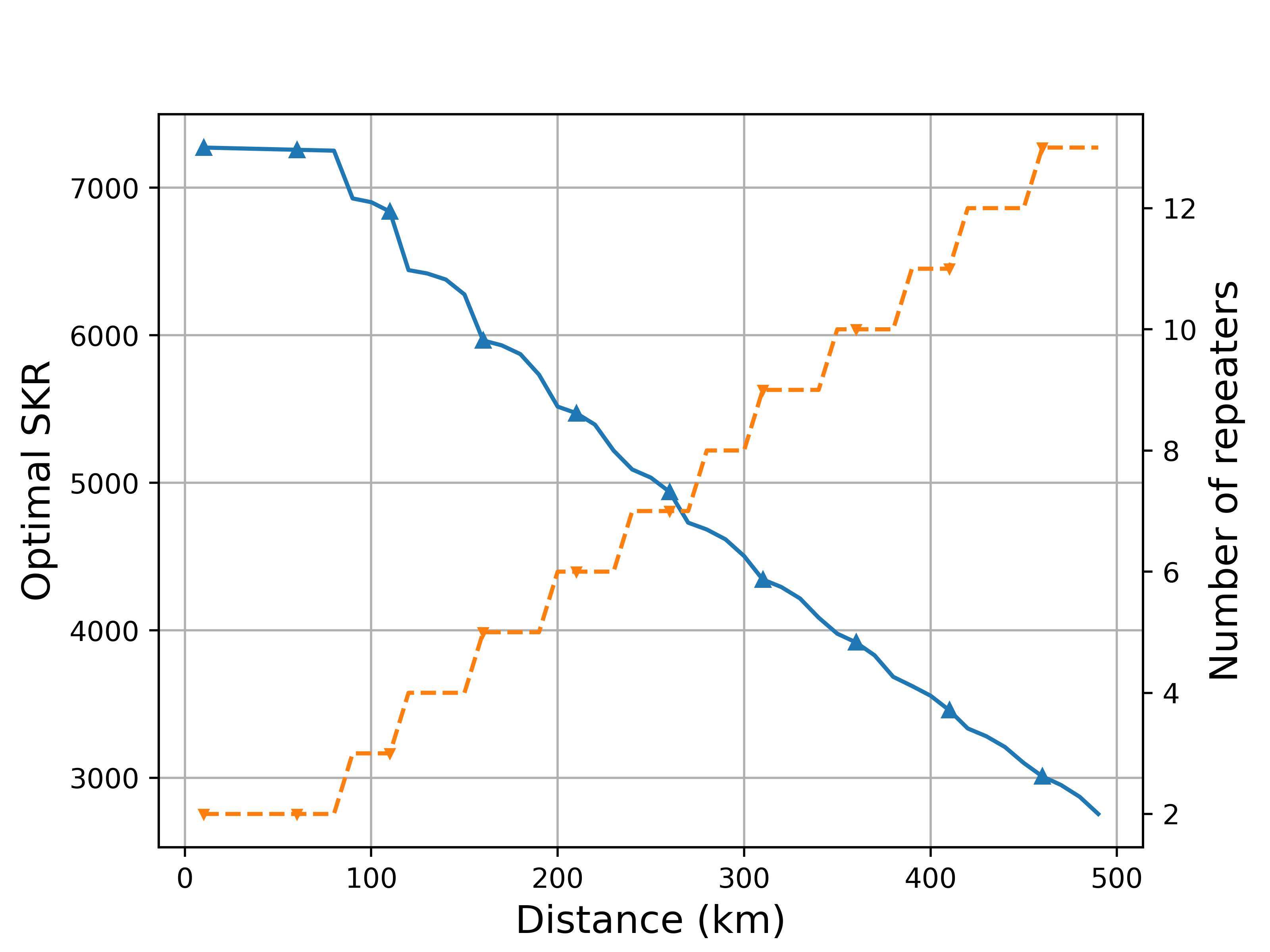}\label{subfig:highLoad}}
    \end{centering}
    \caption{Optimal SKR versus distance with YQF scheduling. We compute the optimal SKR (solid line) with distance, optimizing the number of repeaters (dashed line) for $\lambda = 2000$ and $\lambda = 8000$. The optimal number of repeaters depends on $\lambda$. We set $w^{*} = 0.995$, $\alpha^{*} = 0.996$, and $T^{*} = 1$s .}
    \label{fig:optimalSKR}
\end{figure}

We analyze the relationship between SKR and distance reporting our results in Fig.~\ref{fig:optimalSKR}. We compute the maximum SKR for each \ac{e2e} distance optimizing over the number of repeaters. The curves show that: 1) increasing the number of repeaters improves the maximum SKR with distance for both values of $\lambda$ considered, 2) increasing $\lambda$ also increases the optimal number of repeaters that maximize SKR.


\subsection{Decoherence and Gate Infidelities}

\begin{figure}
    \begin{centering}
        \subfloat[$\lambda = 2000$ requests/s]{\includegraphics[width=\linewidth]{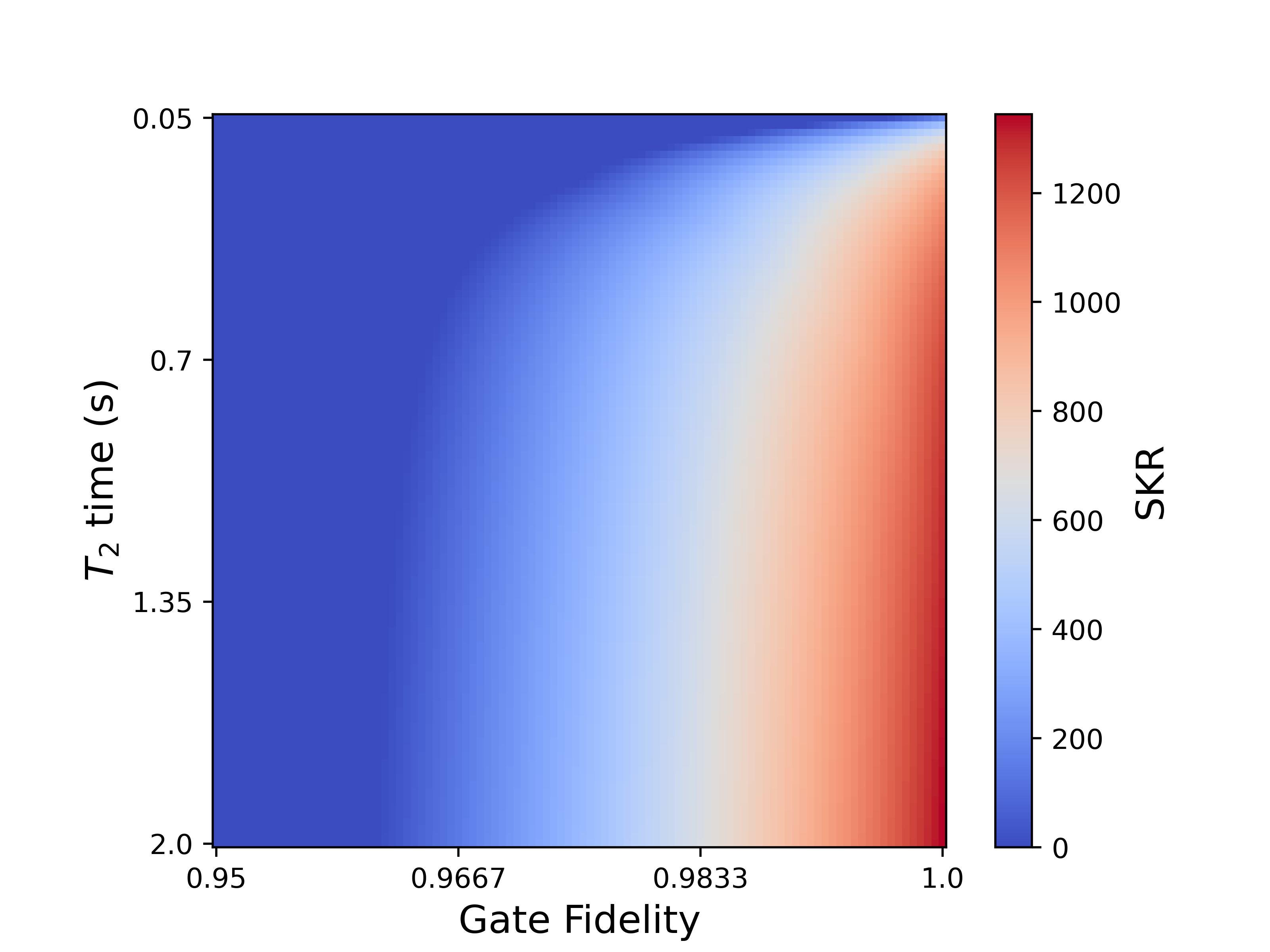} \label{subfig:lowLoad}} \hfill
        \subfloat[$\lambda = 8000$  requests/s] {\includegraphics[width=\linewidth]{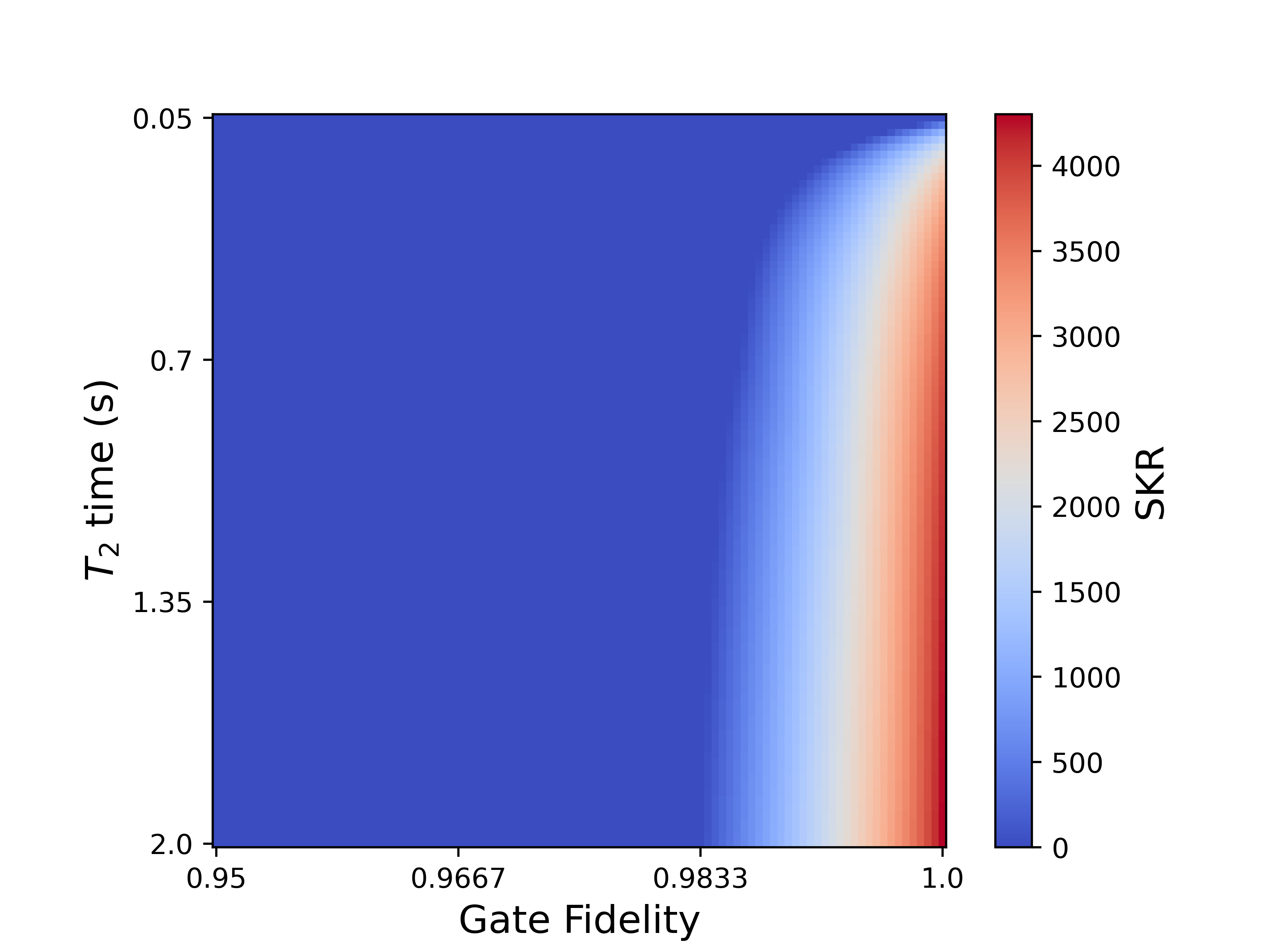}\label{subfig:highLoad}}
    \end{centering}
    \caption{Maximum SKR versus $T_2$ time and gate fidelity for homogeneous repeater chains with YQF scheduling. We maximizing SKR by optimizing it with respect to the number of repeaters in the chain. The distance between end-nodes is 500 km. $w^{*} = 0.995$, for all $j\in\{0,\ldots, n-1\}$.}
    \label{fig:heatmap}
\end{figure}

We now evaluate the impact of decoherence and gate fidelities on the average \ac{e2e} SKR. We report our results in Fig.~\ref{fig:heatmap}. We fix the \ac{e2e} distance to be 500 km. For each combination of memory decoherence time and gate fidelity, we find the number of repeaters that maximize the SKR at $\lambda = 2000$ and $\lambda = 8000$. We optimize over repeater configurations by searching from 2 to 100 repeaters over 500 km. Our results show that, in the parameter regime considered, improvements in gate fidelities have higher benefits than increasing memory coherence times. In particular, we see diminishing returns with improvements in the coherence times. Moreover, this behavior is more nuanced for larger $\lambda$, since the optimal number of repeaters increase.

%% file: src/conclusion.tex
\section{Conclusion}\label{sec:conclusion}
In this work, we built an analytical model for quantum repeater chains with sequential entanglement swapping. We derived closed-form expressions for the expected entanglement fidelity delivered to the end-nodes under the following imperfections: (i) initial Werner states for link-level entangled pairs, (ii) dephasing and depolarizing noise models for memories decoherence, and (iii) noisy quantum operations. We provided closed-form expressions for the one-shot scenario. Moreover, we used queuing theory to model the scenario where end-nodes continuously generate requests. We derived closed-form solutions assuming that requests follow a Poisson arrival process and inter-arrival times for \ac{LLEG} follow an exponential distribution.

We applied the closed-form expressions to evaluate the performances of homogeneous repeater chains. Our results showed that the optimal number of repeaters for a given distance depends on the chain operation regime in terms of arrival rates and hardware parameters. In addition, the results provide insight into operational boundaries for gate fidelity and coherence time for homogeneous repeater chains. Our analysis is easily extended to different parameter regimes due to the generality of our analytical results.

Investigating the effects of finite buffers, multiplexing, and entanglement distillation~\cite{bennett1996mixed} into fidelity and secret-key rate are interesting directions for future work. Standard queuing theory models can be extended to account for finite buffers and multiplexing, although distillation may require non-standard techniques.
In addition, modeling one-way quantum repeater chains with queuing theory is also a promising direction for future work.

\input{src/ack}

%% file: src/ack.tex
{\em Acknowledgments}---This research was supported in part by the NSF grant CNS-1955744, NSF- ERC Center for Quantum Networks grant EEC-1941583, and DOE  Grant AK0000000018297.

%% file: main.bbl
\begin{thebibliography}{10}

\bibitem{buhrman2003distributed}
Harry Buhrman and Hein R{\"o}hrig.
\newblock Distributed quantum computing.
\newblock In {\em International Symposium on Mathematical Foundations of Computer Science}, pages 1--20. Springer, 2003.

\bibitem{zhuang2018distributed}
Quntao Zhuang, Zheshen Zhang, and Jeffrey~H. Shapiro.
\newblock Distributed quantum sensing using continuous-variable multipartite entanglement.
\newblock {\em Phys. Rev. A}, 97:032329, Mar 2018.

\bibitem{vazirani2019qkd}
Umesh Vazirani and Thomas Vidick.
\newblock Fully device independent quantum key distribution.
\newblock {\em Commun. ACM}, 62(4):133, mar 2019.

\bibitem{nielsen2010quantum}
Michael~A Nielsen and Isaac~L Chuang.
\newblock {\em Quantum computation and quantum information}.
\newblock Cambridge university press, 2010.

\bibitem{awschalom2021development}
David Awschalom, Karl~K Berggren, Hannes Bernien, Sunil Bhave, Lincoln~D Carr, Paul Davids, Sophia~E Economou, Dirk Englund, Andrei Faraon, Martin Fejer, et~al.
\newblock Development of quantum interconnects (quics) for next-generation information technologies.
\newblock {\em PRX Quantum}, 2(1):017002, 2021.

\bibitem{craddock2024automated}
Alexander~N. Craddock, Anne Lazenby, Gabriel~Bello Portmann, Rourke Sekelsky, Mael Flament, and Mehdi Namazi.
\newblock Automated distribution of high-rate, high-fidelity polarization entangled photons using deployed metropolitan fibers, 2024.

\bibitem{Preskill2018quantumcomputingin}
John Preskill.
\newblock Quantum {C}omputing in the {NISQ} era and beyond.
\newblock {\em {Quantum}}, 2:79, August 2018.

\bibitem{patil2021multiplexing}
Ashlesha Patil, Joshua~I. Jacobson, Emily Van~Milligen, Don Towsley, and Saikat Guha.
\newblock Distance-independent entanglement generation in a quantum network using space-time multiplexed greenberger–horne–zeilinger (ghz) measurements.
\newblock In {\em 2021 IEEE International Conference on Quantum Computing and Engineering (QCE)}, pages 334--345, 2021.

\bibitem{muralidharan2016optimal}
Sreraman Muralidharan, Linshu Li, Jungsang Kim, Norbert L{\"u}tkenhaus, Mikhail~D Lukin, and Liang Jiang.
\newblock Optimal architectures for long distance quantum communication.
\newblock {\em Scientific reports}, 6(1):20463, 2016.

\bibitem{bennet2014quantum}
Charles~H. Bennett and Gilles Brassard.
\newblock Quantum cryptography: Public key distribution and coin tossing.
\newblock {\em Theoretical Computer Science}, 560:7--11, 2014.

\bibitem{vardoyan2023quantum}
Gayane Vardoyan and Stephanie Wehner.
\newblock Quantum network utility maximization.
\newblock In {\em 2023 IEEE International Conference on Quantum Computing and Engineering (QCE)}, volume~1, pages 1238--1248. IEEE, 2023.

\bibitem{van2024hardware}
Janice van Dam, Guus Avis, Tzula~B Propp, Francisco~Ferreira da~Silva, Joshua~A Slater, Tracy~E Northup, and Stephanie Wehner.
\newblock Hardware requirements for trapped-ion based verifiable blind quantum computing with a measurement-only client.
\newblock {\em arXiv preprint arXiv:2403.02656}, 2024.

\bibitem{lu2024connectionless}
Zirui Xiao, Jian Li, Kaiping Xue, Zhonghui Li, Nenghai Yu, Qibin Sun, and Jun Lu.
\newblock A connectionless entanglement distribution protocol design in quantum networks.
\newblock {\em IEEE Network}, 38(1):131--139, 2024.

\bibitem{brand2020efficient}
Sebastiaan Brand, Tim Coopmans, and David Elkouss.
\newblock Efficient computation of the waiting time and fidelity in quantum repeater chains.
\newblock {\em IEEE Journal on Selected Areas in Communications}, 38(3):619--639, 2020.

\bibitem{coopmans2021netsquid}
Tim Coopmans, Robert Knegjens, Axel Dahlberg, David Maier, Loek Nijsten, Julio de~Oliveira~Filho, Martijn Papendrecht, Julian Rabbie, Filip Rozp{\k{e}}dek, Matthew Skrzypczyk, et~al.
\newblock Netsquid, a network simulator for quantum information using discrete events.
\newblock {\em Communications Physics}, 4(1):164, 2021.

\bibitem{kamin2023exact}
Lars Kamin, Evgeny Shchukin, Frank Schmidt, and Peter van Loock.
\newblock Exact rate analysis for quantum repeaters with imperfect memories and entanglement swapping as soon as possible.
\newblock {\em Physical Review Research}, 5(2):023086, 2023.

\bibitem{goodenough2024noise}
Kenneth Goodenough, Tim Coopmans, and Don Towsley.
\newblock On noise in swap asap repeater chains: exact analytics, distributions and tight approximations, 2024.

\bibitem{chandra2022scheduling}
Aparimit Chandra, Wenhan Dai, and Don Towsley.
\newblock Scheduling quantum teleportation with noisy memories.
\newblock In {\em 2022 IEEE International Conference on Quantum Computing and Engineering (QCE)}, pages 437--446, 2022.

\bibitem{wilde2013quantum}
Mark~M Wilde.
\newblock {\em Quantum information theory}.
\newblock Cambridge university press, 2013.

\bibitem{pirandola2017fundamental}
Stefano Pirandola, Riccardo Laurenza, Carlo Ottaviani, and Leonardo Banchi.
\newblock Fundamental limits of repeaterless quantum communications.
\newblock {\em Nature communications}, 8(1):1--15, 2017.

\bibitem{harchol2013performance}
Mor Harchol-Balter.
\newblock {\em Performance modeling and design of computer systems: queueing theory in action}.
\newblock Cambridge University Press, 2013.

\bibitem{wishart1960queuing}
David~MC Wishart.
\newblock Queuing systems in which the discipline is “last-come, first-served”.
\newblock {\em Operations Research}, 8(5):591--599, 1960.

\bibitem{bennett1996mixed}
Charles~H. Bennett, David~P. DiVincenzo, John~A. Smolin, and William~K. Wootters.
\newblock Mixed-state entanglement and quantum error correction.
\newblock {\em Phys. Rev. A}, 54:3824--3851, Nov 1996.

\end{thebibliography}
